\documentclass{article}
\usepackage[T1]{fontenc}
\usepackage[utf8]{inputenc}
\usepackage{ismir,amsmath,cite,url}
\usepackage{graphicx}
\usepackage{color}
\usepackage{lineno}
\usepackage{ascmac}
\usepackage{booktabs}
\usepackage{siunitx}
\usepackage{seqsplit}

\makeatletter
\let\MYcaption\@makecaption
\makeatother
\usepackage{subcaption}
\makeatletter
\let\@makecaption\MYcaption
\makeatother
\usepackage[cachedir=cache,frozencache]{minted}
\makeatletter
\AtBeginEnvironment{minted}{\dontdofcolorbox}
\def\dontdofcolorbox{\renewcommand\fcolorbox[4][]{##4}}
\makeatother

\let\isCameraReady

\ifdefined\isCameraReady
\else
    \linenumbers
\fi

\title{FruitsMusic: A Real-World Corpus of Japanese Idol-Group Songs}

\ifdefined\isCameraReady
    \multauthor
    {Hitoshi Suda$^1$\hspace{0.3cm}Shunsuke Yoshida$^2$\hspace{0.3cm}Tomohiko Nakamura$^1$\hspace{0.3cm}Satoru Fukayama$^1$\hspace{0.3cm}Jun Ogata$^1$}
    {
        $^1$ National Institute of Advanced Industrial Science and Technology (AIST), Tokyo, Japan\\
        $^2$ The University of Tokyo, Tokyo, Japan\\
        {\tt\small \{suda.h,shunsuke.yoshida,tomohiko-nakamura,s.fukayama,jun.ogata\}@aist.go.jp}\vspace*{-10pt}
    }
    \def\authorname{H. Suda, S. Yoshida, T. Nakamura, S. Fukayama, and J. Ogata}
\else
    \threeauthors
      {First Author} {Affiliation1 \\ {\tt author1@ismir.edu}}
      {Second Author} {\bf Retain these fake authors in\\\bf submission to preserve the formatting}
      {Third Author} {Affiliation3 \\ {\tt author3@ismir.edu}}
    \def\authorname{F. Author, S. Author, and T. Author}
\fi

\usepackage[bookmarks=false,pdfauthor={\authorname},pdfsubject={\papersubject},hidelinks]{hyperref}

\usepackage{cleveref}
\Crefname{figure}{Figure}{Figures}
\crefname{figure}{Figure}{Figures}
\Crefname{table}{Table}{Tables}
\crefname{table}{Table}{Tables}
\Crefname{equation}{Eqn}{Eqn}
\crefname{equation}{Eqn}{Eqn}
\Crefname{section}{Section}{Sections}
\crefname{section}{Section}{Sections}
\Crefname{subsection}{Section}{Sections}
\crefname{subsection}{Section}{Sections}

\begin{document}

\maketitle

\begin{abstract}
This study presents FruitsMusic, a metadata corpus of Japanese idol-group songs in the real world, precisely annotated with who sings what and when.
Japanese idol-group songs, vital to Japanese pop culture, feature a unique vocal arrangement style, where songs are divided into several segments, and a specific individual or multiple singers are assigned to each segment.
To enhance singer diarization methods for recognizing such structures, we constructed FruitsMusic as a resource using 40 music videos of Japanese idol groups from YouTube.
The corpus includes detailed annotations, covering songs across various genres, division and assignment styles, and groups ranging from 4 to 9 members.
FruitsMusic also facilitates the development of various music information retrieval techniques, such as lyrics transcription and singer identification, benefiting not only Japanese idol-group songs but also a wide range of songs featuring single or multiple singers from various cultures.
This paper offers a comprehensive overview of FruitsMusic, including its creation methodology and unique characteristics compared to conversational speech.
Additionally, this paper evaluates the efficacy of current methods for singer embedding extraction and diarization in challenging real-world conditions using FruitsMusic.
Furthermore, this paper examines potential improvements in automatic diarization performance through evaluating human performance.
\end{abstract}

\section{Introduction}
In Japanese pop culture, an \textit{idol} is a performer who engages in dancing, singing, and entertaining fans\cite{Galbraith2012-pk}.
In the culture, idols frequently participate in activities, such as concerts and television programs, as members of idol groups.
One of the most renowned contemporary idol groups is AKB48, which has 40 single compact discs (CDs) that are million-sellers, as certified by The Recording Industry Association of Japan\footnote{\url{https://www.riaj.or.jp/f/data/cert/gd_search.html}}.
FRUITS ZIPPER has emerged as another notable group comprising seven girls and being awarded the Best New Artist at the Japan Record Awards 2023, the most prestigious accolade in Japanese music culture\cite{noauthor_2023-aj}.
Not only can fans attend concerts, but they can also interact with the idols at handshaking events (\textit{Akushukai}) or bonus events (\textit{Tokutenkai}), where the fans can forge deep connections with the idols\cite{Nozawa2023-pn}.

Idol-group songs feature several unique characteristics.
One notable characteristic is \textit{song division}, also called \textit{utawari} in Japanese\cite{Suda2022-fy,Okada2013-rh}.
This approach involves a dynamic vocal arrangement where the singing roles shift throughout the song; individual members may take turns singing solo lines, or multiple members may sing together in unison.
In particular, the entire group often sings together in the chorus sections, known as \textit{sabi}.
Song division is chosen intentionally to maximize the charm and attractiveness of each idol and song.
Therefore, the analysis of song division is crucial for understanding the structure and expression of songs, as well as the creators' intentions.

Song division plays a crucial role also in shaping audience participation through chants and shouts, known as \textit{calls} and \textit{mixes}, which are indispensable elements of idol-group concerts\cite{Xie2021-hy}.
Fans spontaneously create these chants and shouts, reflecting the song's structure, musical intensity, and song division, specifically which member is assigned to sing at any given moment.
Furthermore, song division significantly influences music videos and concert recordings produced by idol groups, demonstrating its pivotal role in producing and appreciating idol music content.

As previously described, song division is crucial for understanding and enjoying the musical compositions of idol groups.
To aid fans' comprehension, some idol groups release official charts showing how songs are divided among members.
For Korean pop groups with similar features to Japanese idol groups, several fans create \textit{line distribution} videos.
These videos, widely viewed on platforms like YouTube and TikTok, visualize the structures of song division, facilitating a deeper understanding.
Therefore, developing techniques for recognizing song division will help fans enjoy the music compositions and enhance their interaction with idols.
In addition, such advancements will support creators in promoting idol groups.

The task to estimate song division, i.e., \textit{who sings when}, within a music signal is known as \textit{singer diarization}.
This technique has been inspired by speaker diarization, which identifies \textit{who speaks when} in conversational speech\cite{Thlithi2015-ds,Anguera2012-ac,Park2022-fr}.
The singer diarization technique was initially introduced to analyze folk music and has been adapted for Japanese idol-group songs\cite{Suda2022-fy}.
However, existing research has not examined songs from real-world idol groups but from idol-themed games and anime.
These game and anime songs generally belong to narrower genres, feature simpler song division structures, and have vocals that are easier to distinguish, thanks to the distinctive voice qualities of the voice actors.
Further research indicates that in-the-wild audio signals can improve diarization performance in real-world settings, even with small datasets\cite{Fujita2019-qs,Suda2022-fy}.
Consequently, compiling a dataset featuring songs from real-world idol groups is critical for developing practical applications targeting pop culture.

This study addresses the demand for a practical dataset in music information retrieval (MIR) by constructing a new corpus, FruitsMusic.
This corpus consists of detailed annotations about \textit{who sings what and when} in real-world songs performed by Japanese idol groups from YouTube, enabling the advancement of singer diarization techniques and their assessments.
Beyond singer diarization, FruitsMusic also advances various MIR techniques such as lyrics transcription\cite{Mesaros2010-hf,Gao2022-uj}, emotion classification\cite{Yang2009-pd,Kim2010-ah}, singer identification\cite{Nithin2014-yq,Fujihara2010-yd}, and singer-based music search\cite{Goto2010-mr}, for not only Japanese idol-group songs but also a wide array of musical pieces featuring single or multiple singers from different cultures.
A significant advantage of FruitsMusic is its focus on real idol groups, allowing for evaluations in challenging scenarios and enhancing the applicability of MIR techniques in the real world.
This paper details the structure, development methodology, and unique characteristics of FruitsMusic.
The paper also demonstrates the applications of evaluating existing methods in two MIR tasks, singer embedding extraction and diarization, in real-world scenarios.

\section{Structure and Construction Methodology of FruitsMusic}
In this study, we constructed FruitsMusic (Corpus of \textbf{F}ully \textbf{R}eal-World Pop\textbf{u}lar \textbf{I}dol-group Songs from You\textbf{T}ube Video\textbf{s} for \textbf{Mus}ic \textbf{I}nformation Pro\textbf{c}essing) aimed at developing and evaluating various MIR techniques.
This corpus is a collection of annotations for 163 minutes of music video content on YouTube, detailing \textit{who sings what and when}.
The corpus includes annotations for 40 songs performed by 18 different groups, featuring a total of 122 unique female singers, all approximately 20 years of age.
\ifdefined\isCameraReady
    The corpus is available at \url{https://huggingface.co/datasets/fruits-music/fruits-music}\footnote{This paper has been written based on FruitsMusic version 1.2.0.}.
\else
    The corpus is available at \texttt{[URL will be filled in the camera-ready version]}\footnote{This is available as supplementary material on the review stage.}.
\fi

\subsection{Related Works}
Several corpora derived from YouTube have been constructed across various research fields.
The key advantage of this approach is the utilization of a wide range of real-world video and audio content.

For example, ActivityNet and YouTube-8M are benchmark datasets widely used in video processing\cite{Heilbron2015-vf,Abu-El-Haija2016-nc}.
Similarly, YouTube-ASL, a large-scale American Sign Language corpus, originates from YouTube \cite{Uthus2023-vw}.

In the field of audio processing, several corpora have utilized YouTube videos.
AudioSet, for instance, is widely adopted for recognizing and detecting audio events\cite{Gemmeke2017-op}.
VoxLingua107 covers 6,628 hours of speech across 107 languages and is helpful to language detection \cite{Valk2021-py}.
Further, JTubeSpeech consists of extensive Japanese speech data from YouTube and helps the development of diverse speech processing techniques\cite{Takamichi2021-ac}.
Similarly, YODAS consists of 500,000 hours of speech in over 100 languages and makes multilingual speech processing techniques applicable in the wild\cite{Takamichi2021-ac,Li2023-rp}.
Coco-Nut is another corpus with subjective descriptions of voices, designed for controlling speaker identity based on text prompts\cite{Watanabe2023-cf}.

These prior works underscore the effectiveness of YouTube-based corpora, which we also adopted in this study.
Our corpus focuses especially on accuracy and reliability, which are less emphasized in these prior corpora.
In addition, the video-based nature of FruitsMusic facilitates multimodal processing, such as multimodal diarization\cite{Noulas2011-yq}.
Note that these prior corpora have been curated to protect individual privacy rights by excluding personal information, and FruitsMusic also maintains these ethical standards.

\subsection{Structure of the Corpus}
FruitsMusic includes annotations in JavaScript Object Notation (JSON) format, Rich Transcription Time Marked (RTTM) files for diarization, and text files of lyrics.
\Cref{tab:example} presents an example of JSON and RTTM files.

\begin{table}
  \begin{minipage}{\linewidth}
    \centering
    \begin{screen}
      \begin{minted}[fontsize=\tiny]{json}
{
  "id": "XXm01",
  "youtubeId": "YouTube ID",
  "type": "music_video",
  "singerIds": ["XXs01", "XXs02", "XXs04", "XXs05", "XXs06"],
  "title": "Song Title",
  "songStartsAt": 0,
  "duration": 216128,
  "states": [
    {
      "start": 1869,
      "end": 17233,
      "singers": [0, 1, 2, 3, 4],
      "lyrics": "Lyrics 1",
      "realLyrics": null
    },
    {
      "start": 22543,
      "end": 26930,
      "singers": [1],
      "lyrics": "Lyrics 2",
      "realLyrics": null
    }
  ]
}
      \end{minted}
    \end{screen}
    \subcaption{JSON file}
    \label{tab:example-json}
  \end{minipage}
  \begin{minipage}{\linewidth}
    \centering
    \begin{screen}
      \begin{minted}[fontsize=\tiny]{text}
SPEAKER XXm01 1 1.869 15.364 <NA> <NA> XXs01 <NA> <NA>
SPEAKER XXm01 1 1.869 15.364 <NA> <NA> XXs02 <NA> <NA>
SPEAKER XXm01 1 1.869 15.364 <NA> <NA> XXs04 <NA> <NA>
SPEAKER XXm01 1 1.869 15.364 <NA> <NA> XXs05 <NA> <NA>
SPEAKER XXm01 1 1.869 15.364 <NA> <NA> XXs06 <NA> <NA>
SPEAKER XXm01 1 22.543 4.387 <NA> <NA> XXs02 <NA> <NA>
      \end{minted}
    \end{screen}
    \subcaption{RTTM file}
    \label{tab:example-rttm}
  \end{minipage}
  \caption{An example of JSON and RTTM files.}
  \label{tab:example}
\end{table}

\subsubsection{JSON Files}

The JSON files include the following information:
\begin{itemize}
    \setlength{\itemsep}{0pt}
    \setlength{\parskip}{3pt}
    \item \textbf{Song ID}. This field is formed by combining a two-character idol-group ID, the letter ``m'', and a two-digit ID.
    \item \textbf{Video ID on YouTube}.
    \item \textbf{Type of the video}. This field is either of \texttt{\seqsplit{music\_video}}, \texttt{\seqsplit{middle\_music\_video}}, or \texttt{\seqsplit{dance\_practice}}. The names of these types are derived from traditions in Japanese idol culture.
    \item \textbf{List of singer IDs}. Each ID is formed by combining a two-character idol-group ID, the letter ``s'', and a two-digit ID.
    \item \textbf{Song title}. This field aims at natural language processing (NLP) tasks.
    \item \textbf{Start time and duration of the song}. The videos may contain content beyond songs, such as comments from idols. This information is provided to help filter out such content.
    \item \textbf{Singing states}. This is a list of the start and end times of the segment, the singers assigned to the segment, and the lyrics. The \texttt{lyrics} field contains the official lyrics, which may differ from the actual lyrics sung. In such cases, the \texttt{realLyrics} field is used.
\end{itemize}
The time and duration fields are annotated in milliseconds.

\subsubsection{RTTM Files}
The RTTM format is specially designed for speaker diarization tasks, identifying \textit{who speaks when}\cite{National_Institute_of_Standards_and_Technology_NIST2009-nv}.
\Cref{tab:example-rttm} presents an example of an RTTM file.
Within this format, each line details the start time and duration of the segment, as well as the singer's ID.
For simultaneous singing, the format allocates a separate line to each singer, resulting in multiple lines corresponding to the number of singers.

\subsubsection{Text Files of Lyrics}
Lyrics lines may be duplicated in the JSON files to precisely represent \textit{who sings what and when} (e.g., DRm03).
As a result, extracting lyrics from JSON files is not straightforward.
To support the development and assessment of techniques involving lyrics, such as lyrics transcription, FruitsMusic provides separate text files of lyrics.

\subsection{Subsets}
FruitsMusic is split into Subset A and Subset B.
Subset A is designed mainly for training, and Subset B is for evaluation.
However, both subsets can be arbitrarily used for various purposes.
Subset A contains 32 songs, while Subset B has 8 songs.
To ensure unbiased evaluation, Subset B does not contain any singers from Subset A, and each group in Subset B contributes only one song.
The songs in Subset B were chosen to cover various genres (dance, rock, synthpop, etc.) and division styles.
Also, groups in Subset B are generally less famous than those in Subset A, which helps ensure fairer and less biased human evaluation.

\subsection{Song Selection}
We meticulously selected the songs for FruitsMusic to ensure the corpus's reliability and usefulness.
To achieve accurate annotations, we initially gathered extensive knowledge about the idol groups.
We then used reliable sources, including concert recordings and official announcements, for information.
Additionally, to support applications like singer diarization, each singer has at least one solo section within the database.
Moreover, we assign each singer to only one group in FruitsMusic.
While idols may participate in multiple groups or move between groups in reality, we avoid such complexities in this database.
FruitsMusic focuses solely on contemporary songs released from 2022 onwards to reflect the latest music trends.

\subsection{Rules}\label{sec:rules}
This corpus has been constructed using copyrighted materials.
Users are required to follow the licensing agreement specified in the corpus documentation to protect the rights of creators and idols.
The agreement sets three major rules.
First, the copyrighted content of this corpus, such as lyrics texts, is not intended for appreciation or entertainment.
Second, the corpus cannot be used to develop or enhance generative artificial intelligence (AI) techniques, such as singing voice synthesis, voice conversion, lyrics generation, and music creation.
However, users can utilize the corpus for recognition or information extraction tasks, including lyrics recognition, singer embedding extraction, and assessing the naturalness of lyrics or music.
Third, when citing this corpus in any media, including academic works and presentations, users are required to identify both the groups and the singers using the provided IDs and refrain from using their real names.
If the mention of song names is not essential for the discussion, users are also required to refer to them by their respective IDs.

\section{Comparison with Conversational Dataset}

\begin{table}
  \centering
  \vspace{6pt}
  \begin{tabular}{ccc}
    \toprule
    & CHiME-5 & FruitsMusic \\
    \midrule
    Average audio length & \SI{9031}{s} & \SI{244}{s} \\
    \# Speakers & 4 & 4--9 \\
    Average segment length & \SI{2.11}{s} & \SI{4.44}{s} \\
    Total length per speaker & \SI{1159.6}{s} & \SI{15.9}{s} \\
    \midrule
    Segments without speakers & 22.3\% & 23.9\% \\
    Solo segments & 51.4\% & 42.6\% \\
    Multiple-speaker segments & 26.4\% & 33.5\% \\
    Segments with 3+ speakers & 6.4\% & 26.5\% \\
    \bottomrule
  \end{tabular}
  \caption{Comparison of FruitsMusic with the CHiME-5 dataset \cite{Barker2018-yl}, a conversational speech dataset. The ``Total length per speaker'' row indicates the average total duration per speaker in each audio.}
  \label{tab:chime5-vs-fruitsmusic}
\end{table}

This section compares FruitsMusic with the CHiME-5 dataset, a conversational speech dataset designed for The 5th CHiME Speech Separation and Recognition Challenge\cite{Barker2018-yl}, to explore the differences between conversational speech and songs with song division.
The CHiME-5 dataset contains 20 conversational speech instances, each from four speakers.
\Cref{tab:chime5-vs-fruitsmusic} shows the comparison results, considering all subsets of both CHiME-5 and FruitsMusic.

Initially, the average audio length in FruitsMusic is significantly shorter than in CHiME-5.
Unlike conversational speech, which is not limited by specific length constraints, the duration of songs is tightly controlled by the structure of the musical compositions.
Furthermore, the average total duration of speech segments per speaker in FruitsMusic is extremely shorter than in CHiME-5.
This difference arises from the shorter overall audio length and the larger number of singers in FruitsMusic.
Since solo segments play a key role in capturing singer characteristics, developing singer identification techniques under these challenging conditions, different from conversational speech, is essential for improving singer diarization and other MIR systems for songs featuring multiple singers.

The comparison reveals a noteworthy difference in the frequency of simultaneous speakers between CHiME-5 and FruitsMusic.
In particular, sections featuring 3 or more singers in FruitsMusic are significantly longer than in CHiME-5.
This indicates that the methods that treat overlapping speech as segments with two speakers, often adopted in speaker diarization\cite{Plaquet2023-nm}, cannot be directly applied to singer diarization.
Furthermore, about 60\% of segments with multiple singers feature vocals from only a subset of the entire group.
Hence, the assumption that all singers are present in overlapped segments proves ineffective for singer diarization; it is crucial to accurately and independently determine the vocal activity of each singer.

\section{Application 1: Singer Embeddings}\label{sec:embeddings}
Singer embeddings are multidimensional vectors that capture each singer's unique vocal traits.
In singing information processing, high-quality singer embeddings are crucial for enhancing the performance of tasks involving singers, such as singer identification, voice matching, and singer diarization.
This section evaluates two types of embeddings extracted from song segments by a specific group and discusses the effectiveness of each extraction technique in real-world scenarios.
This section visualizes these embeddings to understand their effectiveness in distinguishing singers and provides a numerical analysis of the clustering performance based on singers.

In our evaluation, we compare two types of singer embeddings.
The first type involves x-vectors, traditional yet effective speaker embeddings derived from deep neural networks (DNNs) for speaker identification\cite{Snyder2018-lk}.
Specifically, we utilize an x-vector extractor \texttt{microsoft/wavlm-base-plus-sv}\footnote{\url{https://huggingface.co/microsoft/wavlm-base-plus-sv}}, which incorporates WavLM, a large-scale pre-trained model based on self-supervised learning\cite{Chen2022-es}.
Second, we evaluated embeddings based on ECAPA-TDNN, an enhanced time delay neural network (TDNN) in x-vector extractors\cite{Desplanques2020-mx}.
ECAPA-TDNN-based embeddings have been proven to show remarkable performance in speaker recognition and diarization\cite{Desplanques2020-mx,Dawalatabad2021-ev,Nguyen2022-uc}.
We used an ECAPA-TDNN model provided by SeechBrain\footnote{\url{https://huggingface.co/speechbrain/spkrec-ecapa-voxceleb}}\cite{Ravanelli2021-zm}.
In addition, this evaluation considers both mixed and vocal signals, with the latter extracted using Demucs, an open-source music source separation tool\cite{Defossez2021-kw,Rouard2023-ee}.
We utilized the \texttt{htdemucs\_ft} model, a fine-tuned version of the Hybrid Transformer Demucs, renowned for its state-of-the-art performance in music source separation.

We focus on the group KF, which comprises seven members and has eight songs, the most available on FruitsMusic.
For this study, we selected segments where a singer performs solo for over 2 seconds.
On average, each singer has 20 segments, totaling approximately 101 seconds of solo performance.

As objective evaluation metrics, $F$ values are calculated to benchmark clustering efficacy\cite{Hotho2005-qv}.
Here, the $F$ value is the harmonic mean of two metrics: purity $P$ and inverse purity $I$.
The $P$ and $I$ are defined as follows:
\begin{align}
  P &= \frac{1}{N} \sum_{i} \max_{j} |C_i \cap S_j|,\,\text{and}\\
  I &= \frac{1}{N} \sum_{j} \max_{i} |C_i \cap S_j|.
\end{align}
In these equations, $C_i$ is the $i$-th cluster, and $S_j$ is the set of the $j$-th singer's samples.
High $F$ values indicate superior performance, with a theoretical maximum of 1.
For this evaluation, spectral clustering\cite{Shi2000-xg} was performed to create 7 clusters, matching the number of singers.
All the embeddings were $L_2$-normalized in advance.

\begin{figure}
  \centering
  \begin{minipage}{.495\linewidth}
    \centering
    \includegraphics[width=\linewidth]{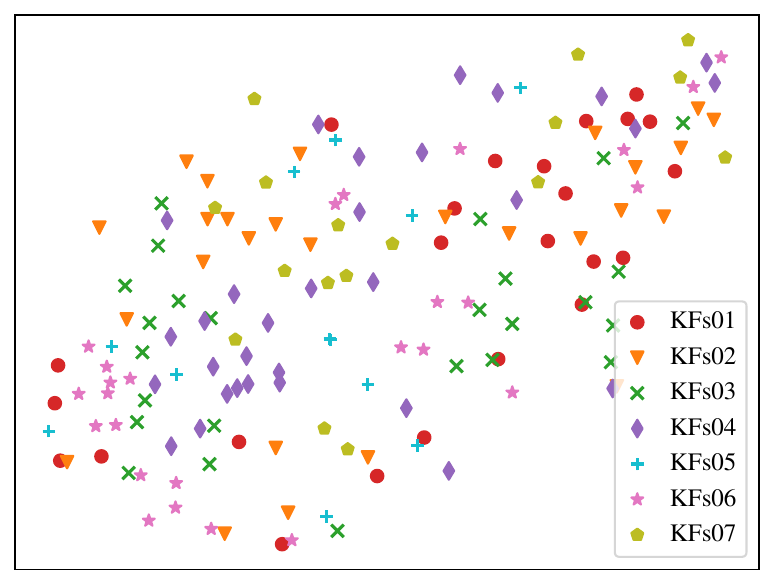}
    \subcaption{x-vector, mixed (0.32)}
  \end{minipage}
  \begin{minipage}{.495\linewidth}
    \centering
    \includegraphics[width=\linewidth]{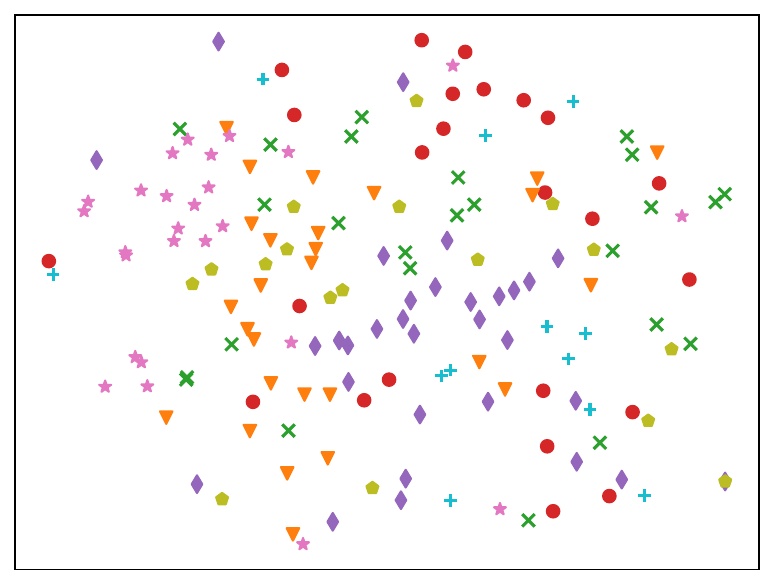}
    \subcaption{x-vector, separated (0.40)}
  \end{minipage}\\[4pt]
  \begin{minipage}{.495\linewidth}
    \centering
    \includegraphics[width=\linewidth]{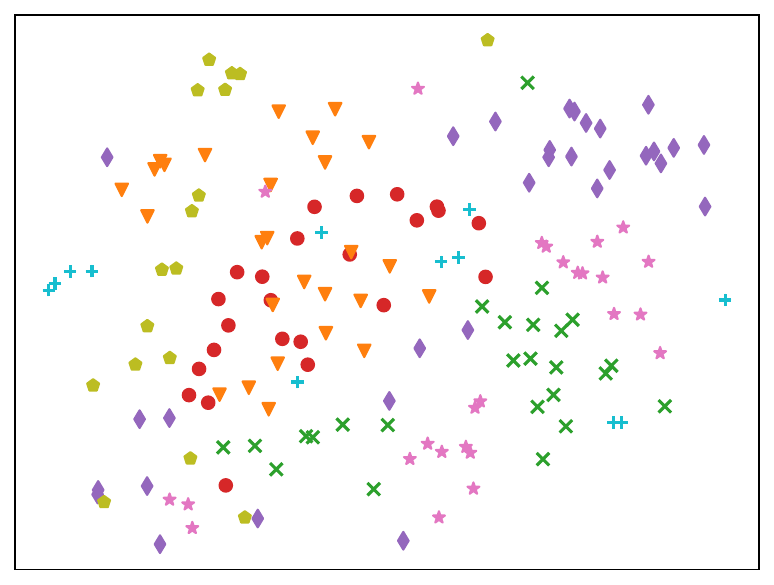}
    \subcaption{ECAPA-TDNN, mixed (0.59)}
  \end{minipage}
  \begin{minipage}{.495\linewidth}
    \centering
    \includegraphics[width=\linewidth]{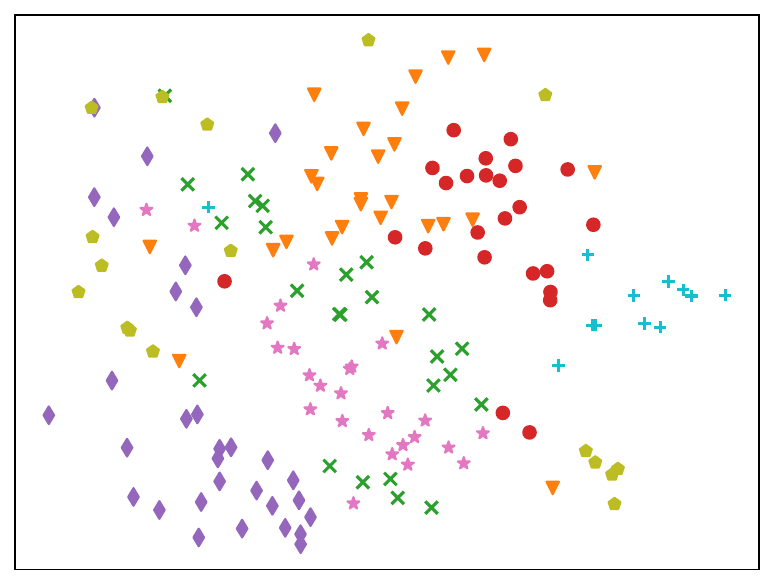}
    \subcaption{ECAPA-TDNN, separated (\textbf{0.64})}
    \label{fig:tsne-ecapa-demucs}
  \end{minipage}

  \caption{$t$-SNE visualizations of singer embeddings from the idol group KF's songs, where each color and shape represents a different singer. Captions detail the extraction methods and whether Demucs was applied. Values in parentheses represent $F$ values, measuring the clustering performance.}
  \label{fig:tsne}
\end{figure}

\Cref{fig:tsne} shows the visualizations of the acquired embeddings by reducing their dimensions into two using $t$-SNE.
The effectiveness of Demucs is confirmed across both extraction methods.
Compared to x-vectors, embeddings derived with ECAPA-TDNN provide more expressive singer representations.
Specifically, \Cref{fig:tsne-ecapa-demucs} reveals that samples from certain singers, specifically KFs01 (represented in red circles) and KFs06 (in pink stars), tend to gather by singer identity.
Hence, ECAPA-TDNN-based embeddings are proved to effectively capture singers' unique identities even from short singing segments.
This shows the advantages of the ECAPA-TDNN methodology over conventional TDNN in x-vector extractors.
However, none of the plots show distinct clusters visibly forming, and the highest $F$ value was only 0.64.
This indicates that tasks like singer diarization and number estimation remain challenging using any embedding extractor evaluated.
Since the ECAPA-TDNN model is trained with speech datasets, fine-tuning it with singing voice datasets will enhance its performance.
Note that the separated vocal signals are distorted; therefore, using datasets with both clean and mixed or separated signals from real-world conditions will be effective.

\section{Application 2: Singer Diarization}
To evaluate the efficacy of FruitsMusic in training singer diarization models, we trained several models with Subset A of FruitsMusic and assessed their performance using Subset B.
In this comparison, the number of singers was not given to the systems.
Furthermore, we engaged a human evaluator to perform the manual diarization of songs in Subset B and discuss the potential advancements in automatic diarization performance.

\subsection{Construction of a Synthesized Dataset}
To improve the diarization performance, we utilized songs from commercial CDs in addition to FruitsMusic.
This dataset consists of 272 songs performed by multiple singers, with a separate recording for each song and singer combination.
For example, if three singers perform song A, each of the three singers has solo recordings: one by singer 1, another by singer 2, and a third by singer 3.
On average, each song features 4.1 singers, resulting in a total of 1126 recordings.
All the songs were sourced from idol-themed games and anime, and the singers were 129 unique female voice actresses.
We executed source separation on all 1126 recordings using Demucs to generate isolated vocal and accompaniment signals.

We generated five song division patterns for each song, capping the number of singers to a maximum of seven.
We applied voice activity detection (VAD) first and randomly assigned singers to each segment.
During the assignment, a single singer was allocated to 60\% of the segments, all singers to 23\%, and random singers to the remaining 17\%.
We mixed the vocal signals based on the generated song division and combined the separated accompaniment with the mixed vocal tracks to create the final mixture.
In these generated songs with song division, singers perform in unison during segments with multiple singers.
The VAD process used \texttt{pyannote/voice-activity-detection}\footnote{\url{https://huggingface.co/pyannote/voice-activity-detection}}.

\subsection{Evaluated Systems}
In this experiment, the following systems are compared.

\subsubsection{SA-EEND with EDA}
The first approach adopted Self-Attentive End-to-End Neural Diarization (SA-EEND)\cite{Fujita2019-qs}.
Since the number of singers for the evaluated signals was unknown, we used enhanced SA-EEND with Encoder-Decoder-based Attractors (EDA)\cite{Horiguchi2020-qe}.
The hyperparameters matched those of the CALLHOME dataset, as specified in the original publication\cite{Horiguchi2020-qe}.
The input signals were downsampled to \SI{8000}{Hz} and were converted to monaural signals.

\subsubsection{\texttt{pyannote.audio}}
The second method used \texttt{pyannote.audio}\footnote{\url{https://github.com/pyannote/pyannote-audio}}, an open-source toolkit for speech processing tasks\cite{Bredin2023-fw,Plaquet2023-nm}.
The diarization workflow is structured as a pipeline process, incorporating PyanNet-based modules.
To conduct this experiment, we fine-tuned the publicly available pre-trained model \texttt{pyannote/speaker-diarization-3.1}\footnote{\url{https://huggingface.co/pyannote/speaker-diarization-3.1}} using the prepared song datasets.
This fine-tuning process adapted the segmentation models and optimized the thresholds for both segmentation and clustering.
The input signals were downsampled to \SI{16000}{Hz} and converted to monaural signals.

\subsubsection{Human Evaluator}
In addition to the automatic diarization approaches, we also engaged a human evaluator to perform manual singer diarization to gauge the achievable performance.
The individual understands Japanese and often listens to Japanese pop music (about 60 hours a month), yet was completely unfamiliar with any of the songs in Subset B of FruitsMusic.
To maintain the experiment's integrity, we presented only the audio signals of the songs without any corresponding videos.
The participant was allowed to use any external tool to aid in the diarization process but was explicitly restricted from searching for the songs on the internet.

\subsection{Experimental Setup}
As a training dataset, Subset A from FruitsMusic was used.
The songs DRm01, KFm01, RGm01, SBm01, and SYm01 were designated for validation.
The remaining songs, excluding three songs featuring nine singers, were allocated for training.
Due to the highly extended training time required for SA-EEND with EDA for songs featuring more than seven singers, three songs with nine singers, VYm02, VYm03, and XSm02, were excluded from the dataset.
The loudness of all songs was normalized to \SI{-14}{LUFS}.

The evaluation metric used was the diarization error rate (DER)\cite{National_Institute_of_Standards_and_Technology_NIST2009-nv}, defined as:
\begin{equation}
  \text{DER} = \frac{\sum_{s=1}^S d_s\left[\max\left(N^{(\mathrm{ref})}_s,N^{(\mathrm{hyp})}_s\right) - N^{(\mathrm{correct})}_s\right]}{\sum_{s=1}^S d_s N^{(\mathrm{ref})}_s}.
\end{equation}
Here, $S$ is the total number of segments, $d_s$ represents the duration of the $s$-th segment, and $N_s^{(\mathrm{ref})}$, $N_s^{(\mathrm{hyp})}$, and $N_s^{(\mathrm{correct})}$ correspond to the number of ground-truth singers, estimated singers, and accurately identified singers in the $s$-th segment, respectively.
According to this definition, DER can exceed 100\%.
The calculation of DER was performed with dscore\footnote{\url{https://github.com/nryant/dscore}}, an open-source tool.
Due to the implementation of dscore, self-overlapped segments, which contain multiple recordings of the same singer, were normalized in the calculation process.
The collar size, the time ignored in DER calculation around segment boundaries, was set to zero.

The model selection criterion was achieving the minimum DER on the validation set.
For each condition, we developed two versions of the system: one trained on mixed signals and another trained on extracted vocal signals.
The vocal signal extraction was performed using the \texttt{htdemucs\_ft} model of Demucs\cite{Defossez2021-kw,Rouard2023-ee}.

\subsection{Results}

\begin{table}
  \centering
  \begin{tabular}{lrr}
    \toprule
    System & Mixed & Separated \\
    \midrule
    \multicolumn{3}{l}{\textbf{SA-EEND with EDA}} \\
    Synthesized only & 99.5\% & 101.3\% \\
    Synthesized + FruitsMusic & 103.2\% & 83.8\% \\
    \midrule
    \multicolumn{3}{l}{\texttt{\textbf{pyannote.audio}}} \\

    Synthesized only & 92.9\% & 69.9\% \\
    Synthesized + FruitsMusic & 91.3\% & 50.3\% \\
    \midrule
    Human & \textbf{22.7\%} & --- \\
    \bottomrule
  \end{tabular}
  \caption{DER for Subset B in FruitsMusic with the several diarization systems.}
  \label{tab:der}
\end{table}

\begin{table}
  \centering
  \begingroup
  \setlength{\tabcolsep}{3.8pt}
  \begin{tabular}{lcccccccc}
    \toprule

    System & BD & BI & JA & JY & MG & QD & SL & TJ \\
    \midrule
    \multicolumn{9}{l}{\textbf{SA-EEND with EDA}} \\
    w/o FruitsMusic & 1 & 3 & 2 & 2 & 4 & 0 & \textbf{6} & 2 \\
    w/ FruitsMusic & 2 & 2 & 2 & 2 & 2 & 2 & 2 & 2 \\
    \midrule
    \multicolumn{9}{l}{\texttt{\textbf{pyannote.audio}}} \\

    w/o FruitsMusic & 3 & 5 & 3 & 3 & 3 & 3 & 3 & 3 \\
    w/ FruitsMusic & 7 & 7 & \textbf{7} & 7 & \textbf{7} & 6 & 7 & 7 \\
    \midrule
    Human & 8 & 6 & 6 & \textbf{5} & \textbf{7} & \textbf{5} & \textbf{6} & \textbf{4} \\
    \midrule
    Ground truth & 9 & 4 & 7 & 5 & 7 & 5 & 6 & 4 \\
    \bottomrule
  \end{tabular}
  \endgroup
  \caption{Estimated total number of singers derived from diarization results. All the systems used separated vocal signals using Demucs. Each column shows a song in Subset B. The suffixes ``m01'' of song IDs are omitted.}
  \label{tab:total-singers}
\end{table}

\Cref{tab:der} shows the DER of all the systems.
The performance of the mixed signal systems is significantly inferior to that of the separated signal systems.
In other words, across evaluated systems, Demucs effectively improved diarization performance; hence, a pipeline system combining source separation and diarization proved more effective than using a single system on mixed signals in the case of this evaluation.
In both approaches, SA-EEND and \texttt{pyannote.audio}, training with FruitsMusic significantly improved the overall performance, particularly for the separated signal systems.
The results suggest that FruitsMusic, despite its smaller size, can significantly enhance diarization performance rather than relying solely on large-scale synthesized datasets.

\Cref{tab:total-singers} shows the estimated number of singers included in the diarization results.
The SA-EEND-based systems struggled to distinguish singers accurately.
This seems due to the difficulties of naive DNN-based methods in distinguishing singer identities, as discussed in \cref{sec:embeddings}.
On the other hand, \texttt{pyannote.audio} demonstrated an almost invariant estimation of the number of singers.
This indicates a potential overfitting to the training datasets, with the most common number of singers in the training set tending to dominate the predictions.

Among the evaluated systems, human performance was remarkably superior to the automatic diarization systems in terms of DER.
Notably, a human evaluator accurately estimated the number of singers in 5 out of 8 songs.
This demonstrates that humans can effectively distinguish individual singers' voices even within mixed music signals.
Therefore, these results proved a significant potential for improving both automatic singer identification and diarization performance.

\section{Conclusion}
This paper presents FruitsMusic, a novel corpus of precise annotations on \textit{who sings what and when} in Japanese idol-group songs.
The song selection and subset creation were meticulously conducted to facilitate unbiased evaluation and ensure usefulness across a wide range of genres, song division styles, and idol groups.
The corpus can be applied to various MIR tasks, such as singer diarization, singer identification, and lyrics transcription.
This paper showcases its applications in evaluating singer embedding extraction and diarization techniques.
The results showed that distinguishing singers from short singing segments remains challenging, despite effective methods in speech processing.
The paper also suggests potential advancements in automatic diarization performance by assessing human performance.
We acknowledge significant existing areas for performance improvement in diverse MIR tasks, and we are confident that FruitsMusic has the potential to advance various techniques among them.

\clearpage
\ifdefined\isCameraReady
    \section{Acknowledgments}
    This work was supported by JSPS KAKENHI Grant Number JP23K20017.
    This paper is based on results obtained from a project, JPNP20006, commissioned by the New Energy and Industrial Technology Development Organization (NEDO).
    This study was supported by the BRIDGE program of the Cabinet Office, Government of Japan.
\fi

\section{Ethics Statement}
The FruitsMusic corpus is derived from YouTube videos of Japanese idol groups and consists of annotations that detail \textit{who sings what and when}.
Hence, FruitsMusic has the potential to help develop singing voice synthesis (SVS) or voice conversion (VC) systems that replicate the voices of actual idols.
Moreover, analysis of the videos within FruitsMusic enables associating the real names of groups and singers with their IDs.
These scenarios raise potential concerns about infringing on the personality rights of the idols.
Additionally, using FruitsMusic to construct or trigger other generative AI techniques, such as lyrics and music generation, may violate the rights of composers and creators.
In light of these considerations, as described in \cref{sec:rules}, FruitsMusic enforces the stringent licensing agreement and requires all users to adhere to it when downloading and using the corpus.

\bibliography{paperpile}

\begin{thebibliography}{10}
\providecommand{\url}[1]{#1}
\csname url@samestyle\endcsname
\providecommand{\newblock}{\relax}
\providecommand{\bibinfo}[2]{#2}
\providecommand{\BIBentrySTDinterwordspacing}{\spaceskip=0pt\relax}
\providecommand{\BIBentryALTinterwordstretchfactor}{4}
\providecommand{\BIBentryALTinterwordspacing}{\spaceskip=\fontdimen2\font plus
\BIBentryALTinterwordstretchfactor\fontdimen3\font minus \fontdimen4\font\relax}
\providecommand{\BIBforeignlanguage}[2]{{%
\expandafter\ifx\csname l@#1\endcsname\relax
\typeout{** WARNING: IEEEtran.bst: No hyphenation pattern has been}%
\typeout{** loaded for the language `#1'. Using the pattern for}%
\typeout{** the default language instead.}%
\else
\language=\csname l@#1\endcsname
\fi
#2}}
\providecommand{\BIBdecl}{\relax}
\BIBdecl

\bibitem{Galbraith2012-pk}
P.~W. Galbraith and J.~G. Karlin, ``Introduction: The mirror of idols and celebrity,'' in \emph{Idols and Celebrity in {J}apanese Media Culture}, P.~W. Galbraith and J.~G. Karlin, Eds., Aug. 2012, pp. 1--32.

\bibitem{noauthor_2023-aj}
``{FRUITS} {ZIPPER} was awarded the {B}est {N}ew {A}rtist at the {J}apan {R}ecord {A}wards 2023 (in {J}apanese),'' \emph{Sankei Shimbun}, Dec. 2023.

\bibitem{Nozawa2023-pn}
S.~Nozawa, ``\BIBforeignlanguage{en}{The concentration booth and the handshaking lane: Ideologies of the phatic},'' \emph{\BIBforeignlanguage{en}{International Journal of the Sociology of Language}}, vol. 2023, no. 284, pp. 15--36, Nov. 2023.

\bibitem{Suda2022-fy}
H.~Suda, D.~Saito, S.~Fukayama, T.~Nakano, and M.~Goto, ``Singer diarization for polyphonic music with unison singing,'' \emph{IEEE/ACM Transactions on Audio, Speech, and Language Processing}, vol.~30, pp. 1531--1545, May 2022.

\bibitem{Okada2013-rh}
Y.~Okada, Ed., \emph{\BIBforeignlanguage{ja}{Living with Idols (in Japanese)}}.\hskip 1em plus 0.5em minus 0.4em\relax Pot Publishing, Jul. 2013.

\bibitem{Xie2021-hy}
W.~Xie, ``Japanese ``idols'' in trans-cultural reception: the case of {AKB48},'' in \emph{The Art of Reception}, J.~Bracker and A.-K. Hubrich, Eds., Mar. 2021, pp. 371--399.

\bibitem{Thlithi2015-ds}
M.~Thlithi, C.~Barras, J.~Pinquier, and T.~Pellegrini, ``Singer diarization: application to ethnomusicological recordings,'' in \emph{Proc. of the 5th International Workshop on Folk Music Analysis ({FMA} 2015)}, Jun. 2015, pp. 124--125.

\bibitem{Anguera2012-ac}
X.~Anguera, S.~Bozonnet, N.~Evans, C.~Fredouille, G.~Friedland, and O.~Vinyals, ``Speaker diarization: A review of recent research,'' \emph{IEEE Transactions on Audio, Speech, and Language Processing}, vol.~20, no.~2, pp. 356--370, Feb. 2012.

\bibitem{Park2022-fr}
T.~J. Park, N.~Kanda, D.~Dimitriadis, K.~J. Han, S.~Watanabe, and S.~Narayanan, ``A review of speaker diarization: Recent advances with deep learning,'' \emph{Computer Speech \& Language}, vol.~72, no.~C, pp. 1--34, Mar. 2022.

\bibitem{Fujita2019-qs}
Y.~Fujita, N.~Kanda, S.~Horiguchi, Y.~Xue, K.~Nagamatsu, and S.~Watanabe, ``End-to-end neural speaker diarization with self-attention,'' in \emph{Proc. of the 2019 {IEEE} Automatic Speech Recognition and Understanding Workshop ({ASRU})}, Dec. 2019, pp. 296--303.

\bibitem{Mesaros2010-hf}
A.~Mesaros and T.~Virtanen, ``\BIBforeignlanguage{en}{Automatic recognition of lyrics in singing},'' \emph{\BIBforeignlanguage{en}{EURASIP Journal on Audio, Speech, and Music Processing}}, vol. 2010, no.~1, pp. 1--11, Feb. 2010.

\bibitem{Gao2022-uj}
X.~Gao, C.~Gupta, and H.~Li, ``Automatic lyrics transcription of polyphonic music with lyrics-chord multi-task learning,'' \emph{IEEE/ACM Transactions on Audio, Speech, and Language Processing}, vol.~30, pp. 2280--2294, Jul. 2022.

\bibitem{Yang2009-pd}
D.~Yang and W.-S. Lee, ``Music emotion identification from lyrics,'' in \emph{Proc. of the 11th {IEEE} International Symposium on Multimedia}, Dec. 2009, pp. 624--629.

\bibitem{Kim2010-ah}
Y.~E. Kim, E.~M. Schmidt, R.~Migneco, B.~G. Morton, P.~Richardson, J.~Scott, J.~A. Speck, and D.~Turnbull, ``Music emotion recognition: A state of the art review,'' in \emph{Proc. of the 11th International Society for Music Information Retrieval Conference ({ISMIR} 2010)}, vol.~86, Aug. 2010, pp. 937--952.

\bibitem{Nithin2014-yq}
C.~Nithin and J.~Cheriyan, ``A novel approach to automatic singer identification in duet recordings with background accompaniments,'' in \emph{Proc. of the 2014 Annual International Conference on Emerging Research Areas: Magnetics, Machines and Drives ({AICERA/iCMMD})}, Jul. 2014, pp. 1--6.

\bibitem{Fujihara2010-yd}
H.~Fujihara, M.~Goto, T.~Kitahara, and H.~G. Okuno, ``A modeling of singing voice robust to accompaniment sounds and its application to singer identification and vocal-timbre-similarity-based music information retrieval,'' \emph{IEEE Transactions on Audio, Speech, and Language Processing}, vol.~18, no.~3, pp. 638--648, Mar. 2010.

\bibitem{Goto2010-mr}
M.~Goto, T.~Saitou, T.~Nakano, and H.~Fujihara, ``Singing information processing based on singing voice modeling,'' in \emph{Proc. of the 2010 {IEEE} International Conference on Acoustics, Speech and Signal Processing ({ICASSP})}, Mar. 2010, pp. 5506--5509.

\bibitem{Heilbron2015-vf}
F.~C. Heilbron, V.~Escorcia, B.~Ghanem, and J.~C. Niebles, ``{ActivityNet}: A large-scale video benchmark for human activity understanding,'' in \emph{2015 {IEEE} Conference on Computer Vision and Pattern Recognition ({CVPR})}, Jun. 2015.

\bibitem{Abu-El-Haija2016-nc}
S.~Abu-El-Haija, N.~Kothari, J.~Lee, P.~Natsev, G.~Toderici, B.~Varadarajan, and S.~Vijayanarasimhan, ``{YouTube-8M}: A large-scale video classification benchmark,'' \emph{arXiv [cs.CV] 1609.08675}, Sep. 2016.

\bibitem{Uthus2023-vw}
D.~Uthus, G.~Tanzer, and M.~Georg, ``{YouTube-ASL}: A large-scale, open-domain american sign language-english parallel corpus,'' in \emph{37th Conference on Neural Information Processing Systems ({NeurIPS} 2023) Track on Datasets and Benchmarks}, Jun. 2023.

\bibitem{Gemmeke2017-op}
J.~F. Gemmeke, D.~P.~W. Ellis, D.~Freedman, A.~Jansen, W.~Lawrence, R.~Channing~Moore, M.~Plakal, and M.~Ritter, ``Audio {S}et: An ontology and human-labeled dataset for audio events,'' in \emph{Proc. of the 2017 {IEEE} International Conference on Acoustics, Speech and Signal Processing ({ICASSP})}, Mar. 2017, pp. 776--780.

\bibitem{Valk2021-py}
J.~Valk and T.~Alum{\"a}e, ``{VoxLingua107}: A dataset for spoken language recognition,'' in \emph{Proc. of the 2021 {IEEE} Spoken Language Technology Workshop ({SLT})}, Jan. 2021, pp. 652--658.

\bibitem{Takamichi2021-ac}
S.~Takamichi, L.~K{\"u}rzinger, T.~Saeki, S.~Shiota, and S.~Watanabe, ``{JTubeSpeech}: corpus of japanese speech collected from {YouTube} for speech recognition and speaker verification,'' \emph{arXiv [cs.SD] 2112.09323}, Dec. 2021.

\bibitem{Li2023-rp}
X.~Li, S.~Takamichi, T.~Saeki, W.~Chen, S.~Shiota, and S.~Watanabe, ``Yodas: {Youtube-Oriented} dataset for audio and speech,'' in \emph{Proc. of the 2023 {IEEE} Automatic Speech Recognition and Understanding Workshop ({ASRU})}, Dec. 2023, pp. 1--8.

\bibitem{Watanabe2023-cf}
A.~Watanabe, S.~Takamichi, Y.~Saito, W.~Nakata, D.~Xin, and H.~Saruwatari, ``{Coco-Nut}: Corpus of {J}apanese utterance and voice characteristics description for prompt-based control,'' in \emph{Proc. of the 2023 {IEEE} Automatic Speech Recognition and Understanding Workshop ({ASRU})}, Sep. 2023.

\bibitem{Noulas2011-yq}
A.~Noulas, G.~Englebienne, and B.~J.~A. Krose, ``\BIBforeignlanguage{en}{Multimodal speaker diarization},'' \emph{\BIBforeignlanguage{en}{IEEE Transactions on Pattern Analysis and Machine Intelligence}}, vol.~34, no.~1, pp. 79--93, Mar. 2011.

\bibitem{National_Institute_of_Standards_and_Technology_NIST2009-nv}
{National Institute of Standards and Technology (NIST)}, ``The 2009 ({RT-09}) rich transcription meeting recognition evaluation plan,'' Tech. Rep., 2009.

\bibitem{Barker2018-yl}
J.~Barker, S.~Watanabe, E.~Vincent, and J.~Trmal, ``The fifth {`CHiME'} speech separation and recognition challenge: Dataset, task and baselines,'' in \emph{Proc. of the Interspeech 2018}, Mar. 2018, pp. 1561--1565.

\bibitem{Plaquet2023-nm}
A.~Plaquet and H.~Bredin, ``Powerset multi-class cross entropy loss for neural speaker diarization,'' in \emph{Proc. of the Interspeech 2023}, 2023, pp. 3222--3226.

\bibitem{Snyder2018-lk}
D.~Snyder, D.~Garcia-Romero, G.~Sell, D.~Povey, and S.~Khudanpur, ``X-vectors: Robust {DNN} embeddings for speaker recognition,'' in \emph{Proc. of the 2018 {IEEE} International Conference on Acoustics, Speech and Signal Processing ({ICASSP})}, Apr. 2018, pp. 5329--5333.

\bibitem{Chen2022-es}
S.~Chen, C.~Wang, Z.~Chen, Y.~Wu, S.~Liu, Z.~Chen, J.~Li, N.~Kanda, T.~Yoshioka, X.~Xiao, J.~Wu, L.~Zhou, S.~Ren, Y.~Qian, Y.~Qian, J.~Wu, M.~Zeng, X.~Yu, and F.~Wei, ``{WavLM}: Large-scale self-supervised pre-training for full stack speech processing,'' \emph{IEEE Journal of Selected Topics in Signal Processing}, vol.~16, no.~6, pp. 1505--1518, Oct. 2022.

\bibitem{Desplanques2020-mx}
B.~Desplanques, J.~Thienpondt, and K.~Demuynck, ``{ECAPA-TDNN}: Emphasized channel attention, propagation and aggregation in {TDNN} based speaker verification,'' in \emph{Proc. of the Interspeech 2020}, ISCA, Oct. 2020.

\bibitem{Dawalatabad2021-ev}
N.~Dawalatabad, M.~Ravanelli, F.~Grondin, J.~Thienpondt, B.~Desplanques, and H.~Na, ``{ECAPA-TDNN} embeddings for speaker diarization,'' \emph{arXiv [eess.AS] 2104.01466}, Apr. 2021.

\bibitem{Nguyen2022-uc}
T.~L. Nguyen, B.~T. Ta, D.~Van~Hai, T.~A.~X. Tran, and N.~M. Le, ``Speaker diarization for {V}ietnamese conversations using deep neural network embeddings,'' in \emph{Proc. of the 2022 {IEEE} Ninth International Conference on Communications and Electronics ({ICCE})}, Jul. 2022, pp. 300--305.

\bibitem{Ravanelli2021-zm}
M.~Ravanelli, T.~Parcollet, P.~Plantinga, A.~Rouhe, S.~Cornell, L.~Lugosch, C.~Subakan, N.~Dawalatabad, A.~Heba, J.~Zhong, J.-C. Chou, S.-L. Yeh, S.-W. Fu, C.-F. Liao, E.~Rastorgueva, F.~Grondin, W.~Aris, H.~Na, Y.~Gao, R.~D. Mori, and Y.~Bengio, ``{{SpeechBrain}}: A general-purpose speech toolkit,'' \emph{arXiv [eess.AS] 2106.04624}, Jun. 2021.

\bibitem{Defossez2021-kw}
A.~D{\'e}fossez, ``Hybrid spectrogram and waveform source separation,'' in \emph{Proc. of the {ISMIR} 2021 Music Demixing Workshop}, Nov. 2021, pp. 1--11.

\bibitem{Rouard2023-ee}
S.~Rouard, F.~Massa, and A.~D{\'e}fossez, ``Hybrid transformers for music source separation,'' in \emph{Proc. of the 2023 {IEEE} International Conference on Acoustics, Speech and Signal Processing ({ICASSP})}, Jun. 2023, pp. 1--5.

\bibitem{Hotho2005-qv}
A.~Hotho, A.~N{\"u}rnberger, and G.~Paa{\ss}, ``A brief survey of text mining,'' \emph{Journal for Language Technology and Computational Linguistics}, vol.~20, no.~1, pp. 19--62, Jul. 2005.

\bibitem{Shi2000-xg}
J.~Shi and J.~Malik, ``\BIBforeignlanguage{en}{Normalized cuts and image segmentation},'' \emph{\BIBforeignlanguage{en}{IEEE transactions on pattern analysis and machine intelligence}}, vol.~22, no.~8, pp. 888--905, 2000.

\bibitem{Horiguchi2020-qe}
S.~Horiguchi, Y.~Fujita, S.~Watanabe, Y.~Xue, and K.~Nagamatsu, ``End-to-end speaker diarization for an unknown number of speakers with encoder-decoder based attractors,'' in \emph{Proc. of the Interspeech 2020}, May 2020, pp. 269--273.

\bibitem{Bredin2023-fw}
H.~Bredin, ``{Pyannote.Audio} 2.1 speaker diarization pipeline: Principle, benchmark, and recipe,'' in \emph{Proc. of the Interspeech 2023}, Aug. 2023.

\end{thebibliography}

\end{document}